\documentclass[twocolumn,preprint]{aastex63}
\usepackage{amsmath,amstext}
\usepackage[T1]{fontenc}
\usepackage{apjfonts} 
\usepackage{natbib}
\usepackage{longtable}
\usepackage{comment}

\newcommand{\hst}{\textit{HST}}
\newcommand{\hstlong}{\textit{Hubble Space Telescope}}
\newcommand{\jwst}{\textit{JWST}}
\newcommand{\jwstlong}{\textit{James Webb Space Telescope}}

\newcommand{\cloudy}{\texttt{Cloudy}}

\newcommand{\Ha}{\hbox{{\rm H}$\alpha$}}
\newcommand{\ha}{\hbox{{\rm H}$\alpha$}}
\newcommand{\Hb}{\hbox{{\rm H}$\beta$}}
\newcommand{\hb}{\hbox{{\rm H}$\beta$}}
\newcommand{\Hg}{\hbox{{\rm H}$\gamma$}}

\newcommand{\neiii}{\hbox{[\ion{Ne}{3}]}}
\newcommand{\xneiii}{\hbox{[\ion{Ne}{3}] $\lambda 3870$}}
\newcommand{\nev}{\hbox{[\ion{Ne}{5}]}}
\newcommand{\xnev}{\hbox{[\ion{Ne}{5}] $\lambda 3347,3427$}}
\newcommand{\xneva}{\hbox{[\ion{Ne}{5}] $\lambda 3427$}}
\newcommand{\oii}{\hbox{[\ion{O}{2}]}}
\newcommand{\xoii}{\hbox{[\ion{O}{2}] $\lambda\lambda 3727,3729$}}
\newcommand{\oiii}{\hbox{[\ion{O}{3}]}}
\newcommand{\xoiii}{\hbox{[\ion{O}{3}] $\lambda\lambda 4960,5008$}}
\newcommand{\xoiiia}{\hbox{[\ion{O}{3}]$\lambda 5008$}}

\newcommand{\heii}{\hbox{\ion{He}{2}}}
\newcommand{\sii}{\hbox{[\ion{S}{2}]}}
\newcommand{\nii}{\hbox{[\ion{N}{2}]}}
\newcommand{\civ}{\hbox{\ion{C}{4}}}

\newcommand{\ariv}{\hbox{[\ion{Ar}{4}]}}

\newcommand{\mbh}{\hbox{$M_\mathrm{BH}$}}
\newcommand{\msun}{\hbox{$M_\odot$}}
\newcommand{\msol}{\hbox{$M_\odot$}}

\newcommand{\zsun}{\hbox{$Z_\odot$}}

\newcommand{\snr}{\hbox{S/N}}

\begin{document}


\title{\large \bf Using [\ion{Ne}{5}]/[\ion{Ne}{3}] to Understand the Nature of Extreme-Ionization Galaxies}

\correspondingauthor{Nikko J. Cleri}
\email{cleri@tamu.edu}

\author[0000-0001-7151-009X]{Nikko J. Cleri}
\affiliation{Department of Physics and Astronomy, Texas A\&M University, College Station, TX, 77843-4242 USA}
\affiliation{George P.\ and Cynthia Woods Mitchell Institute for Fundamental Physics and Astronomy, Texas A\&M University, College Station, TX, 77843-4242 USA}

\author[0000-0002-4606-4240]{Grace M. Olivier}
\affiliation{Department of Physics and Astronomy, Texas A\&M University, College Station, TX, 77843-4242 USA}
\affiliation{George P.\ and Cynthia Woods Mitchell Institute for Fundamental Physics and Astronomy, Texas A\&M University, College Station, TX, 77843-4242 USA}

\author[0000-0001-6251-4988]{Taylor A. Hutchison}
\altaffiliation{NASA Postdoctoral Fellow}
\affiliation{Astrophysics Science Division, NASA Goddard Space Flight Center, 8800 Greenbelt Rd, Greenbelt, MD 20771, USA}

\author[0000-0001-7503-8482]{Casey Papovich}
\affiliation{Department of Physics and Astronomy, Texas A\&M University, College Station, TX, 77843-4242 USA}
\affiliation{George P.\ and Cynthia Woods Mitchell Institute for Fundamental Physics and Astronomy, Texas A\&M University, College Station, TX, 77843-4242 USA}

\author[0000-0002-1410-0470]{Jonathan R. Trump}
\affiliation{Department of Physics, 196 Auditorium Road, Unit 3046, University of Connecticut, Storrs, CT 06269}

\author[0000-0001-5758-1000]{Ricardo O. Amor\'{i}n}
\affiliation{Instituto de Investigaci\'{o}n Multidisciplinar en Ciencia y Tecnolog\'{i}a, Universidad de La Serena, Raul Bitr\'{a}n 1305, La Serena 2204000, Chile}
\affiliation{Departamento de Astronom\'{i}a, Universidad de La Serena, Av. Juan Cisternas 1200 Norte, La Serena 1720236, Chile}

\author[0000-0001-8534-7502]{Bren E. Backhaus}
\affiliation{Department of Physics, 196 Auditorium Road, Unit 3046, University of Connecticut, Storrs, CT 06269}

\author[0000-0002-4153-053X]{Danielle A. Berg}
\affiliation{Department of Astronomy, The University of Texas, Austin, Texas, 78712 USA} 

\author[0000-0003-0531-5450]{Vital Fern\'{a}ndez}
\affiliation{Instituto de Investigaci\'{o}n Multidisciplinar en Ciencia y Tecnolog\'{i}a, Universidad de La Serena, Raul Bitr\'{a}n 1305, La Serena 2204000, Chile}

\author[0000-0001-8519-1130]{Steven L. Finkelstein}
\affiliation{Department of Astronomy, The University of Texas, Austin, Texas, 78712 USA} 

\author[0000-0001-7201-5066]{Seiji Fujimoto}
\affiliation{Cosmic Dawn Center (DAWN), Jagtvej 128, DK2200 Copenhagen N, Denmark}
\affiliation{Niels Bohr Institute, University of Copenhagen, Lyngbyvej 2, DK2100 Copenhagen \O, Denmark}

\author[0000-0002-3301-3321]{Michaela Hirschmann}
\affiliation{Institute of Physics, Laboratory of Galaxy Evolution, EPFL, Observatoire de Sauverny, 1290 Versoix, Switzerland}

\author[0000-0001-9187-3605]{Jeyhan S. Kartaltepe}
\affiliation{Laboratory for Multiwavelength Astrophysics, School of Physics and Astronomy, Rochester Institute of Technology, 84 Lomb Memorial Drive, Rochester, NY 14623, USA}

\author[0000-0002-8360-3880]{Dale D. Kocevski}
\affiliation{Department of Physics and Astronomy, Colby College, Waterville, ME 04901, USA}

\author[0000-0002-6386-7299]{Raymond C. Simons}
\affiliation{Department of Physics, 196 Auditorium Road, Unit 3046, University of Connecticut, Storrs, CT 06269}

\author[0000-0003-3903-6935]{Stephen M.~Wilkins} %
\affiliation{Astronomy Centre, University of Sussex, Falmer, Brighton BN1 9QH, UK}
\affiliation{Institute of Space Sciences and Astronomy, University of Malta, Msida MSD 2080, Malta}

\author[0000-0003-3466-035X]{{L. Y. Aaron} {Yung}}
\altaffiliation{NASA Postdoctoral Fellow}
\affiliation{Astrophysics Science Division, NASA Goddard Space Flight Center, 8800 Greenbelt Rd, Greenbelt, MD 20771, USA}

\begin{abstract}
Spectroscopic studies of extreme-ionization galaxies (EIGs) are critical to our understanding of exotic systems throughout cosmic time. These EIGs exhibit spectral features requiring >54.42 eV photons: the energy needed to fully ionize helium into He$^{2+}$ and emit \heii\ recombination lines. Spectroscopic studies of EIGs can probe exotic stellar populations or accretion onto intermediate-mass black holes ($\sim 10^{2}-10^{5}$~\msol), which are possibly key contributors to the reionization of the Universe. To facilitate the use of EIGs as probes of high-ionization systems, we focus on ratios constructed from several rest-frame UV/optical emission-lines: \xoiiia, \hb, \xneiii, \xoii, and \xneva. These lines probe the relative intensity at energies of 35.12, 13.62, 40.96, 13.62, and 97.12 eV, respectively,  covering a wider range of ionization than traced by other common rest-frame UV/optical techniques. We use ratios of these lines (\nev/\neiii\ $\equiv$ Ne53, \oiii/\hb, and \neiii/\oii) which are nearby in wavelength, mitigating effects of dust attenuation and uncertainties in flux calibration.  We make predictions from photoionization models constructed from \cloudy\ that use a broad range of stellar populations and black hole accretion models to explore the sensitivity of these line ratios to changes in the ionizing spectrum. We compare our models to observations from the \hstlong\ and \jwstlong\ of galaxies with strong high-ionization emission lines at $z\sim0$, $z\sim2$, and $5<z<8.5$. We show that the Ne53 ratio can separate galaxies with ionization from ``normal'' stellar populations from those with AGN and even ``exotic'' Population III models. We introduce new selection methods to identify galaxies with photoionization driven by Population III stars or intermediate-mass black hole accretion disks that could be identified in upcoming high-redshift spectroscopic surveys.   
\end{abstract}

\section{Introduction} \label{sec:intro}
Emission-line spectroscopy provides a wealth of information about the physical conditions of a galaxy. From a rest-frame ultraviolet (UV)/optical spectrum of a galaxy, we can discern chemical abundances and metallicities \citep[e.g.,][]{Lequeux1979}, star formation rates \citep[e.g.,][]{Kennicutt1998,Kennicutt2012}, nebular dust attenuation estimates \citep[e.g.,][]{Groves2012,Buat2002}, temperatures and densities of the interstellar medium (ISM) \citep[e.g.,][]{Kewley2019b,Dopita2000,Maiolino2019}, and contributions of active galactic nuclei (AGN) to the emission \citep[e.g.,][]{Baldwin1981,Veilleux1987,Trump2015}. Much of the knowledge of these physical properties of a galaxy's ionized gas is derived from bright Balmer lines of hydrogen (\Ha\ and \Hb), along with metal lines from oxygen (\oii\ $\lambda\lambda$3727,3729 and \oiii\ $\lambda\lambda$4960,5008), sulfur (\sii\ $\lambda\lambda$ 6718,6733) and nitrogen (\nii\ $\lambda$ 6585)\footnote{Vacuum wavelengths}, while P-cygni stellar wind features characterize the massive star population. 

Historically, studies of emission-line galaxies use ratios of emission-line flux  with small wavelength separations as a way to minimize uncertainties. Commonly-used emission-line ratios in the ultraviolet (UV) and optical such as $\oiii \lambda 5008/\Hb$, $\neiii \lambda 3870/\oii \lambda 3728$,  $\nii\lambda6585/\Ha$, $\sii\lambda6718/\Ha$, are relatively insensitive to dust attenuation and spectral calibrations, and as such are very useful tracers of ISM and AGN narrow-line region (NLR) conditions \citep[e.g.,][]{Baldwin1981,Veilleux1987}. 

These rest-frame UV and optical emission-line ratios are tools for understanding how H-ionizing photons (Lyman continuum, LyC, photons of $\lambda$ < 912 \AA) escaped from high-redshift galaxies. As such, this is paramount at $z > 6$ as this radiation is the most likely candidate for reionizing the Universe \citep[e.g.,][]{Finkelstein2015}. Currently, few studies have successfully employed these UV and optical lines at $z > 6$ during this so-called epoch of reionization (EoR), thus the physical nature of the ionizing spectra and physical properties of the galaxies responsible remain poorly understood. With the advent of the \jwstlong\ (\jwst) era, we now have unprecedented access to spectroscopic observations of these early-Universe systems. The Near-Infrared Spectrograph (\jwst/NIRSpec) coverage spans 0.6-5.3 \micron, allowing detection of rest-frame UV and optical emission lines well into the EoR, with spectral resolutions ($R\sim1000$ or $R\sim2700$ depending on the observing mode) easily capable of resolving important emission features for these studies.  

The chemically young and highly energetic stellar populations in galaxies during the EoR are expected to be key contributors to the hydrogen reionization of the Universe \citep[e.g.,][]{Berg2016,Berg2019,Berg2021,Finkelstein2015,Curtis-Lake2023,Trump2023,Brinchmann2023,Zackrisson2011,Trussler2022}. These objects show prominent high-ionization ($\gtrsim$35 eV creation potential, \citealt{Berg2021}) emission lines, suggesting that they are characterized by hard radiation fields \citep{Smit2014,Stark2016,Trump2023,Brinchmann2023,katz2023}.

Previous studies of high-ionization emission-line galaxies (ELGs) have focused primarily on spectral features in the \cite{Berg2021} ``high-ionization regime'', including \civ\ $\lambda\lambda$1548,1551, \neiii\ $\lambda$3870, \oiii\ $\lambda\lambda$4960,5008, \heii $\lambda$1640, \heii\ $\lambda$4687, and \ariv\ $\lambda\lambda$4712,4741  \citep[e.g.,][]{Berg2016,Berg2019,Berg2021,Olivier2022,Levesque2014,Masters2014,Zeimann2015,Backhaus2022a,Atek2011,vanderWel2011b,Maseda2013,Maseda2014,Tang2019,Tang2021,Senchyna2017,Senchyna2020,Rigby2015,Kehrig2015,Kehrig2018,Kehrig2021,Amorin2015,Amorin2017,Perez-Montero2021}. Several studies have already suggested that chemically-evolved Population I and II stars are unable to account for the strength of emission in the highest-energy lines, and instead they require additional sources of (hard) ionizing spectra \citep[e.g.,][]{Steidel2014,Olivier2022}.

Such systems, which we dub ``extreme-ionization galaxies'' (EIGs), exhibit spectral features which require radiation hard enough to fully ionize helium (>54.42 eV). These galaxies often require harder ionization than produced from  ``normal'' stellar populations  to produce such high-energy emission features. Such ionizing sources may be more exotic than ``normal'' stellar populations, including accreting massive black holes, supernovae, Population III stars, Wolf-Rayet stars, stripped stars in binaries, high-mass X-ray binaries (HMXBs), hot, low-mass evolved stars (HOLMES, \citealt{Flores-Farjado2011}). Features seen in spectra of EIGs may include lines from helium (e.g., \heii\ $\lambda$1640, \heii\ $\lambda$4687), nitrogen (e.g, \ion{N}{5} $\lambda\lambda$1240,1244), oxygen (e.g., \ion{O}{4}] $\lambda\lambda$1401,1405), neon (e.g., [\ion{Ne}{4}] $\lambda$ 2423, \nev\ $\lambda\lambda$3347,3427), iron (e.g., [\ion{Fe}{10}] $\lambda$6375), and many others. These conditions appear indicative of rapidly accreting black holes or low-metallicity, recently formed stellar populations. We expect these conditions to be more apparent at higher redshifts, particularly in the EoR. Early work with \jwst\ suggests that many EoR galaxies have metallicities of $\sim 5-10$\% \zsun\ \citep[e.g.,][]{Trump2023,Brinchmann2023}, and have also suggested the existence of active black hole accretion in the very early Universe  \citep[e.g.,][]{Brinchmann2023,Larson2023,Maiolino2023}. In light of these recent results, it is prudent to consider the conditions that could produce EIGs at these early epochs. 

In this work, we focus on emission in EIGs from quadruply-ionized neon through \xneva\footnote{The near-UV/optical \nev\ is a doublet with lines at 3427~\AA\ and 3347~\AA.  However, the 3347~\AA\ line is weaker (with a typical ratio of 2.73:1 for \xneva/\nev\ $\lambda $3347) and the \nev\ $\lambda$ 3347 line can be blended with other nearby lines  (see Section \ref{subsec:ne53} for more discussion). We therefore focus on \nev\ $\lambda$3427 in this work.} (vacuum wavelength) and its ratios with other bright UV/optical lines.

The ionization energy needed to produce Ne$^{4+}$ (97.11 eV) is extremely high compared to most other strong UV/optical emission lines; \nev\ emission requires energies nearly triple that of \oiii\ (35.12 eV), more than double that of \neiii\ (40.96 eV), and nearly double that of \heii\ (54.42 eV). This property places \nev\ well above the lower bound of the \cite{Berg2021} ``very high'' ionization zone, and as such probes a parameter space of the ISM not traced by other strong UV/optical lines\footnote{The  $\gtrsim$100 eV energy regime also ionizes both hydrogen and helium to produce HI and \heii\ photons; however, the minimum energy required to produce \heii\ emission (54.42 eV) and HI emission (13.6 eV) is a factor of $\sim$2--7 lower than that of \nev\ (97.11 eV).}

The production of high-ionization emission lines such as \nev\ requires an extremely hard photoionizing source. Previous works have attributed \nev\ production to photoionization from active galactic nuclei (AGN), stellar continuum from an extremely hot ionizing spectrum including Wolf-Rayet stars, or energetic radiative shocks from supernovae \citep{Zeimann2015,Backhaus2022a,Gilli2010,Mignoli2013,Izotov2012,Cleri2023}.

AGN, produced from accretion onto black holes (BHs), are clear candidates for production of photons needed to produce high-ionization emission lines.\footnote{Here we use ``Active Galactic Nuclei'' (AGN) to indicate radiation from any accreting black hole.  This includes both AGN associated with supermassive black holes ( $\mbh  \gtrsim 10^6~M_\odot$) and AGN arising from accretion onto more modest ``intermediate'' mass black holes (IMBHs) with $M_\mathrm{BH} \lesssim 5~\msun$, see e.g., \citet{Cann2018}.}   Previous works have studied \nev\ and its correlations with X-ray luminosities in local Seyferts and low-redshift ($z<1.5$) quasi-stellar objects (QSOs) \citep{Gilli2010,Mignoli2013}.  Accretion onto intermediate-mass black holes (IMBHs) has remained an open field of study, though models predict that the accretion disks around IMBHs ($\log(\mbh/\msun) \lesssim 5$) produce harder radiation fields than their supermassive counterparts \citep{Done2012}. These lower-mass accretion disk models are predicted to emit photons hard enough to produce very-high-ionization (>54.42 eV) emission features \citep{Cann2018}. 

Around the peak of cosmic star formation rate density and AGN activity at $z\sim 1-2$, studies have found that \nev\ emission is consistent with photoionization from AGN \citep{Cleri2023,Backhaus2022a}. \cite{Cleri2023} finds that \nev-emitting galaxies at these redshifts are five times more likely to be X-ray confirmed AGN than galaxies without \nev-detections, and that a majority (88\%) of their objects are consistent with AGN emission in optical \citealt{Baldwin1981} (BPT)-like emission-line ratio diagnostics. 

Some studies of local ($z\sim0$) low-mass ($\log M_*/\msun \lesssim8$), star-forming (SF) galaxies have explained \nev\ production through energetic supernova shocks. \cite{Izotov2012} finds five oxygen-poor blue compact dwarf (BCDs) galaxies with \nev\ emission which have \nev/\heii\ flux ratios reproducible by radiative shock models with shock velocities in the 300-500 km s$^{-1}$ range and shock ionizing contributions that are $\sim10\%$ of that from the stellar continuum ionization. However, this modeling cannot conclusively rule out that this $\sim10\%$ contribution of the \nev\ emission comes from AGN. These studies have primarily focused on low-mass galaxies (BCDs in the case of \citealt{Izotov2012}), so it is unclear if shocks can account for \nev\ emission in more massive galaxies that dominate the peak of cosmic star formation rate density. Alternatively, \citealt{Olivier2022} has attributed local \nev\ production to young ($\sim$5 Myr) extremely metal-poor ($Z\lesssim 0.1 Z_\odot$) bursts of star formation plus an additional very hot ionizing source, modeled by an 80-100 kK blackbody (in excess of what is available in ``normal'' Population I or II stellar populations).

Another intriguing engine for the energetic photons required to produce \nev\ is the exotic stellar populations in the very early Universe. Population III (Pop III) stars were likely very massive \citep[>50 \msol,][]{Zackrisson2011}, with mass upper limits possibly even as great as 1000 \msol\ \citep{Bromm1999,Nakamura2002,Tan2004,Greif2006,Ohkubo2009}. Such chemically young populations likely initiated the metal enrichment of the succeeding generations of ``normal'' Population I and II stars in the later Universe \citep{Heger2002}. While truly primordial gas would not show metal emission, the chemical enrichment of the ISM of early galaxies occurs rapidly given the short lifetimes of Pop III stars \citep{Finlator2015,Finlator2016,Finlator2018}. However, the short lifetimes of these Pop III stars introduces another complication; observation of high-ionization metal emission lines from the newly-enriched ISM requires very recent ($\lesssim 1$ Myr) star formation for such short-lived stars. 

The spectral energy distributions (SEDs) of these Pop III stars are most likely to peak in the extreme-UV (<912 \AA), around the energies needed to fully ionize helium at 54.42 eV \citep{Zackrisson2011,Schaerer2002,Trussler2022}. Stellar population synthesis codes predict that the SEDs of these Pop III stars may produce photons at even higher energies, even into to very soft X-ray regime \citep[e.g. see the  Yggdrasil models of ][]{Zackrisson2011,Trussler2022}. This would allow for production of ultra high-ionization emission lines like \nev\ to be powered solely from these primordial stars provided the nebula ionized by these stars contains \textit{some} elements heavier than He.

We can alleviate the potential degeneracies between Pop III stars and other high-ionization sources (e.g., AGN, supernova shocks, Wolf-Rayet stars, high-mass X-ray binaries) with deeper information from the rest-frame UV/optical spectra. The UV continuum can be used to derive the metallicity of the stellar populations, and the gas-phase metallicity can be derived from emission features. Other ionizing engines such as AGN, supernova shocks, and Wolf-Rayet stars can produce broader emission features than standard HII regions, and X-ray observations can rule out high-mass X-ray binaries \citep[e.g.,][]{Senchyna2020}.

For objects which are not deeply observed across all wavelengths, separating photoionization from Pop III stars and AGN or other sources is often difficult \citep[e.g.,][]{Katz2022b}. Several works have offered potential solutions in the form of emission line ratios of high-ionization to low-ionization lines (such as \heii/\hb), in conjunction with derived gas-phase metallicities \citep[e.g.,][]{Schaerer2002,Raiter2010,Inoue2011,Katz2022b,Trussler2022}. 

In this work, we offer a potential diagnostic for photoionization from stellar populations or black hole accretion disks using a combination of rest-frame UV/optical emission line ratios (\oiii/\hb, \neiii/\oii, \nev/\neiii).  These line ratios are closely spaced in wavelength, and can be observed out to $z\sim 9$ in upcoming spectroscopic surveys at high-redshifts with \jwst. 

The outline of this work is as follows: Section \ref{sec:models} describes our photoionization models. Section \ref{sec:data} describes our comparison samples. Section \ref{sec:sf_agn} compares UV/optical emission-line ratios as diagnostics of AGN activity. Section \ref{sec:discussion} discusses the implications of our results. Section \ref{sec:conclusions} summarizes the results of this work and discusses future studies of high-ionization galaxies with \jwst.

\section{Photoionization Models}\label{sec:models}
To explore the physical conditions of EIGs throughout cosmic time, we employ photoionization modeling of local stellar populations, black hole accretion disks, and Pop III stars. For the following analysis, we use \cloudy\ version C17.01 \citep{Ferland2017}. \cloudy\ is a photoionization simulation code designed to self-consistently model physical conditions in astrophysical clouds to predict thermal, ionization, and chemical structure of the cloud and predict its observed spectrum. 

Our models assume a hydrogen density of $10^2$ cm$^{-3}$, with the \cloudy\ default \cite{Grevesse2010} solar abundance ratios and Orion dust grains for the initial gas phase and dust abundances. The emissivity of \nev/\neiii\ is relatively insensitive to fluctuations in density (see Appendix B of \citealt{Cleri2023}). Unless otherwise stated, we assume a plane-parallel gas geometry for all models. 

We run all of our models across a grid of ionization parameters\footnote{This is the initial ionization parameter \cloudy\ assumes at the incident face of the cloud.} from $-4<\log U<-1$ in steps of 0.25, where $U=q/c$ is the dimensionless ionization parameter. \cite{Kewley2019b} defines the dimensionless ionization parameter as 
\begin{align}\label{eq:U}
    U &\equiv \frac{1}{c}\frac{\Phi}{n_H}
\end{align}
where $\Phi$ is the ionizing photon flux and $n_H$ is the hydrogen density.

The details of the individual models used in this work are described in the following subsections.

\subsection{``Normal'' Stellar Population Models}
The stellar population models used in this work are from the Binary Population and Spectral Synthesis (BPASS; v2.2.1) single-burst binary-formation library \citep{Stanway2018}. The BPASS-only models use an IMF with a \citealt{Salpeter1955} slope ($\alpha=-2.35$) for all masses, and an upper mass limit of 100\msun. Unless otherwise stated, we use a single-burst stellar population with an age of 3 Myr, with stellar metallicities in a grid of  $Z=0.00005,~0.05,~0.10,~0.20,~0.50,~1.0~Z_\odot$. We note that  $Z=0.05~Z_\odot$ gas-phase metallicity is consistent with extremely metal-poor dwarf galaxies ($Z<0.1 Z_\odot$) in the local Universe, and recent \jwst\ spectra have not derived lower gas-phase metallicities than 5\% solar \citep{Olivier2022,Berg2019,Berg2021,Trump2023,Brinchmann2023}. We also use a composite stellar population model that adds to BPASS models a hot (80 kK) blackbody needed to reproduce emission from very-high-ionization lines such as \heii\ and \civ\ from \citealt{Olivier2022} (see Section \ref{subsec:data:j10j14}). 

\subsection{``Exotic'' Population III Star Models}
To test exotic stellar populations potentially found in the early Universe, we also employ models of Pop III stars from the Yggdrasil population synthesis code \citep{Zackrisson2011}. We use the Pop III.1 model with a zero-metallicity population with an extremely top heavy IMF (50-500 \msol) and a \cite{Salpeter1955} slope ($\alpha=-2.35$) for all masses, with a single stellar population (SSP) from \cite{Schaerer2002}. We also use the Yggdrasil Pop III.2 model, which is also a zero-metallicity stellar population but with a more moderately top-heavy IMF: log-normal with characteristic mass $M_c=10\msol$, dispersion $\sigma=1\msol$, extending from 1-500 \msun\ \citep{Raiter2010,Tumlinson2006}. 

For these models, the stellar population is assumed to have zero metallicity, but we employ a (slightly) enriched gas-phase metallicity of $Z=0.05Z_\odot$. For the very first population of stars, the gas-phase metallicity would be effectively null; this would yield no metal emission features, i.e., only hydrogen and helium emission. We assume that there must be some chemical enrichment in the ISM from primordial supernovae or stellar mass loss events for metals to be present. This 5\% solar gas-phase metallicity is consistent with extremely metal-poor dwarf galaxies in the local Universe, the most likely well-studied analogs to these high-redshift systems \citep[e.g.,][]{Berg2016,Berg2019,Berg2021,Olivier2022}. 

\subsection{Black Hole Accretion Models}

Accretion onto BHs offers another source of high energy photons that are capable of \nev\ production. We tested both theoretical models for accretion onto black holes, as well as empirical models for active galactic nuclei (AGN).  

The theoretical models for black hole accretion are based on SEDs from \cite{Done2012}. These models include a blackbody disk as well as a Compton upscattering component from the disk to recreate the ``soft X-ray excess'' often observed in AGN in the local Universe \citep[e.g.,][]{Done2012,Gierlinski2004,Walter1993}. The final component to this model is a power-law tail to model the high energy emission which is the result of a second Compton upscattering that takes place in the optically thin corona above the disk. The \cite{Done2012} models are available from \texttt{XSPEC} \citep{Arnaud1996} using the \textsc{optxagnf} command. We produce SEDs of black holes with masses from 10$^3$--10$^8$~M$_\odot$ and keep the parameters as the default parameters for this model except for the electron temperature associated with the soft Comptonization component which we set to 0.1~keV instead of the default 0.2~keV as this is more similar to the models used by \cite{Cann2018}. 

Given the intricacies of accurately modeling black hole accretion disk physics, the models used in this work are relatively simplistic \citep{Adhikari2016,Mitchell2023}. They employ a single cloud to represent the broad-line and narrow-line regions, among other simplifications \citep{Cann2018}. More sophisticated models are an active area of work (e.g, Mckaig et al. in prep.). Potential improvements to these models include separate BLR and NLR physics, as well as other updates.  These updated models are not likely to have a significant impact on the parameter space of interest in this work (i.e., $\lesssim$100eV emission lines), but will include significant improvements at higher energies (Mckaig \& Satyapal, private communication). 

We derive the gas-phase metallicities for these black hole accretion disk models from the mass-metallicity relation of \cite{Papovich2022}. We derive respective stellar masses for the galaxies surrounding these black hole accretion disks using the \citealt{Kormendy2013} scaling relations. The \citealt{Papovich2022} mass-metallicity relation is calibrated for more massive ($\gtrsim10^9\msun$) ELGs around the peak of cosmic star formation rate density and AGN activity at $z\sim1-2$, so we extrapolate the relation for the $\mbh\leq10^5\msun$, and set a minimum gas-phase metallicity for our lowest black hole mass models of $Z=0.05Z_\odot$.


To probe a representative of the AGN parameter space with a well-studied local active galaxy, we also include the SED of the nearby Seyfert 1 galaxy NGC 5548 \citep{Mehdipour2015}. NGC 5548 has a multiwavelength SED from near-infrared to the hard (200 keV) X-ray. The UV to near-IR continuum of NGC 5548 is consistent with being composed of a single Comptonized disk component, with no evidence of an additional purely-thermal disk component or additional component of reprocessing from the disk. Notably, NGC 5548 also exhibits a ``soft X-ray excess'', which is correlated with the production of photons required to ionize neon to produce \nev. NGC 5548 has a central engine powered by a $6.5\times10^7$\msun\ supermassive black hole \citep{Bentz2007}. 

Several works suggest other empirical SED shapes for low-redshift Seyferts \citep{Binette1989,Clavel1990} or high-redshift quasars \citep{Zheng1997,Korista1997,VandenBerk2001,Richards2006,Fan2006}, yet these models are similar in shape to the \citealt{Done2012} SEDs. Ultimately, the nature of AGN SEDs remains an open question, especially for the low-mass and low-luminosity regimes. Lower-mass accreting black holes ($\lesssim10^5$\msun) may be key contributors to the high-ionization photon production of the early Universe, yet remain poorly understood. 

\subsection{SEDs and \cloudy\ Output Spectra}
\begin{figure*}[t]
\epsscale{1.1}
\plotone{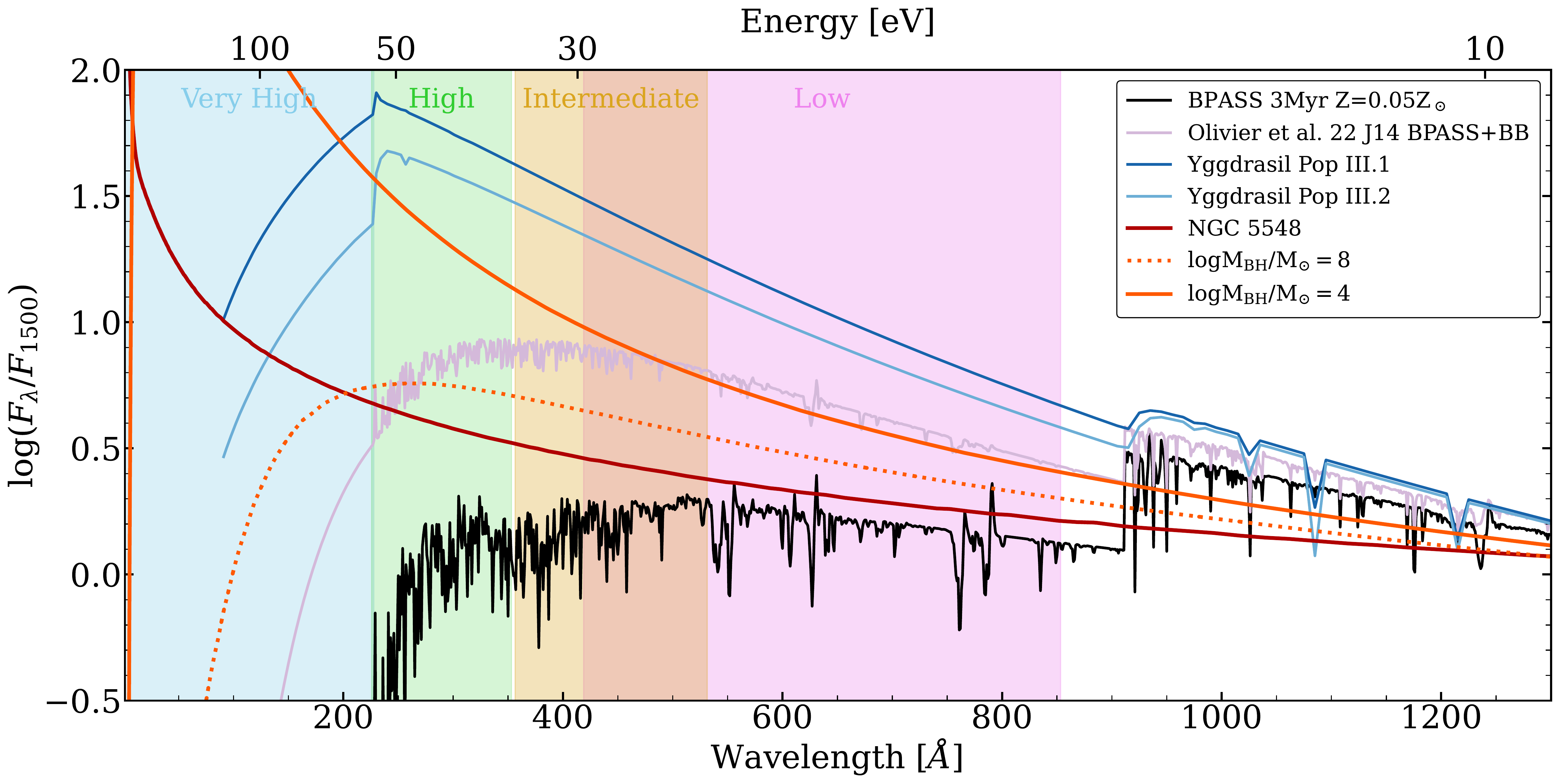}
\caption{Example input SEDs of the various models used in our photoionization modeling. We show a pure BPASS 5\% solar 3 Myr stellar population model in black, the \cite{Olivier2022} BPASS+blackbody model for J141851 in purple, the black hole accretion disk models in orange ($\log\mbh/\msun=8$ dotted, $\log\mbh/\msun=4$ solid), NGC 5548 in red, and the Yggdrasil Pop III.1 and III.2 models in dark and light blue, respectively. All models are normalized in $F_\lambda$ at 1500 \AA. The shaded regions mark the four zones of ionization from \cite{Berg2021} (see Section \ref{sec:models}). 
\label{fig:sed}}
\end{figure*}

Our photoionization models for black hole accretion and stellar populations produce a wide range of emission at different energies. Figure \ref{fig:sed} shows these input SEDs used in our photoionization models.  These include: (1) a 3 Myr BPASS stellar population with continuum and gas-phase metallicity 0.05$Z_\odot$; (2) a model that includes the 3 Myr BPASS stellar population with an additional 80~kK black body (needed to reproduce line ratios seen in the $z\sim0$ metal-poor dwarf galaxy J141851;  \citealt{Olivier2022},  see Section \ref{subsec:data:j10j14} for details),(3) the \citealt{Done2012} black hole accretion disk models for black hole masses ranging from $\log\mbh/\msol=4$ to $\log\mbh/\msol=8$; (4) the \citealt{Mehdipour2015} NGC 5548 SED, and (5) the Yggdrasil Pop III.1 and III.2 SEDs. All of the models shown in Figure \ref{fig:sed} are normalized at 1500 \AA. We mark the four ionization zones of \cite{Berg2021}: 
\begin{enumerate}
    \item Low Ionization: Energy needed to produce N$^+$ (14.53-29.60 eV)
    \item Intermediate Ionization: Energy needed to produce S$^{2+}$ (23.33-34.79 eV)
    \item High Ionization: Energy needed to produce O$^{2+}$ (35.11-54.93 eV)
    \item Very High Ionization: Energy needed to produce He$^{2+}$ (>54.42 eV)
\end{enumerate}

\begin{figure*}[t]
\epsscale{1.1}
\plotone{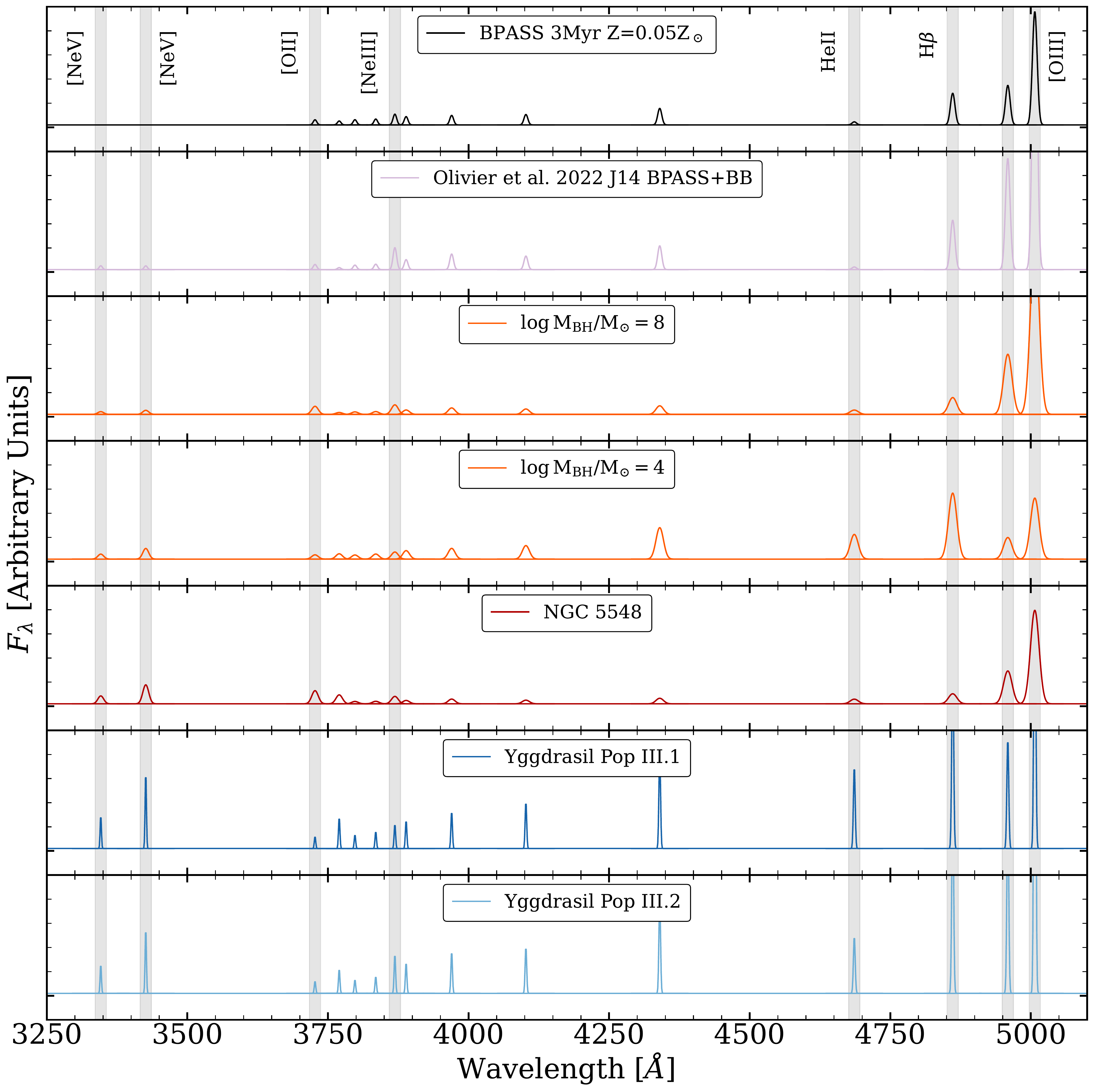}
\caption{ Example \cloudy\ output spectra of the various models used in this work. We show a pure BPASS 5\% \zsun\ 3 Myr stellar population model in black, the \cite{Olivier2022} BPASS+blackbody model for J141851 in purple, the black hole accretion disk models in orange ($\log\mbh/\msun=8$ and $\log\mbh/\msun=4$), NGC 5548 in red, and the Yggdrasil Pop III.1 and III.2 models in dark and light blue, respectively. We note several lines of interest: $\nev~\lambda\lambda$3347,3427, $\oii ~\lambda\lambda$3727,3729, $\neiii ~\lambda$3870, $\heii~\lambda 4687$, \hb,  $\oiii ~\lambda\lambda$4960,5008. For visualization purposes, we smooth the \cloudy\ spectra into Gaussians for each emission line (see Section \ref{sec:models} for details).
\label{fig:output_spectra}}
\end{figure*}

Figure \ref{fig:output_spectra} shows the rest-frame NUV/optical \cloudy\ output intrinsic spectra from the respective ionizing SEDs in Figure \ref{fig:sed}. We mark several lines of interest: \xnev, \xoii, \xneiii, $\heii~\lambda 4687$, \hb, and \xoiii. We note that there is a large apparent difference between the spectra at the wavelengths of the \nev\ lines. There is clear emission of \nev\ in the AGN, NGC 5548, Pop III.1, and Pop III.2 spectra, but minimal \nev\ in the BPASS 3 Myr or \cite{Olivier2022} BPASS+blackbody spectra. For visualization purposes, we smooth the Cloudy emission lines into Gaussians. We assume wind speeds of 250 km/s for the BPASS models and the \citealt{Olivier2022} BPASS+blackbody models (see \citealt{Olivier2022} for details), 460 km/s for the black hole accretion disk models and NGC 5548 \citep{Peterson2013}. For the Yggdrasil Pop III models, we assume minimal 100 km/s winds; winds are metal driven, so true first-population stars will have zero wind \citep[e.g.,][]{Schaerer2002,Kudritzki2000,Baraffe2000}.  The default spectral resolution of \cloudy\ is $R=300$, so we note that there will be blending issues with other lines in the output spectra; as such, this Gaussian smoothing should be used only for visualization purposes. 

\section{Data} \label{sec:data}
We compare our models to galaxies at $z\sim0$ from \cite{Berg2021} and \cite{Olivier2022}, $z\sim1.6$ from \cite{Cleri2023}, and $5<z<8.5$ from \cite{Trump2023}, where for the latter we have made new measurements of \nev\ as part of this work. We briefly describe these observations in the following subsections. 

\subsection{Observations at $z\sim0$} \label{subsec:data:j10j14}
We compare our models with two galaxies in the nearby Universe which exhibit exceptional extreme emission-line behavior. These objects have been studied in detail in \cite{Berg2016,Berg2019,Berg2021,Olivier2022}: 

\begin{itemize}
    \item J104457 has the largest \civ\ $\lambda\lambda 1548, 1550$ \AA\ equivalent width (-6.71 \AA, -2.83 \AA\footnote{Negative equivalent widths indicate emission features.}) measured in the local Universe. 
    \item J141851 has the largest \heii\ $\lambda 1640$ \AA\ equivalent width (-2.82 \AA) measured in the local Universe.
\end{itemize}

J104457 and J141851 have far-UV spectra taken with the Cosmic Origins Spectrograph (COS) on \hst\ (\hst\ Proposal ID 15465) and near-UV/Optical spectra from the Multi-Object Double Spectrograph (MODS) on the Large Binocular Telescope (LBT). They are both extremely metal-poor dwarf galaxies with derived gas-phase metallicities of 5.8\% and 8.7\% solar, respectively, and low stellar masses of $\log M_*/M_\odot$ = 6.80 and 6.63, respectively. 

\subsubsection{Stellar Population + Blackbody Models}
We also employ photoionization models used by \cite{Olivier2022} to reproduce the emission line strengths in two galaxies, J104457 and J141851.  These models add (very hot) blackbodies  to BPASS stellar populations in order to reproduce the emission of high-ionization spectral features such as \civ, and \heii. \citet{Olivier2022} matches physical conditions of the galaxies (densities, abundances, etc.) and the full suite of low- to very-high-ionization lines in the rest-frame far-UV to optical to constrain the full shape of the ionizing spectra.  They find that the models which best reproduce the very-high-ionization emission in their galaxy J141851 (see Section \ref{subsec:data:j10j14}) are those with stellar contributions from BPASS contributing 45\% of the total luminosity, along with an injected 80 kK blackbody contributing the remaining 55\% of the total luminosity (see discussion in the next paragraph). These models use BPASS stellar populations with an IMF with low-mass (<0.5\msun) slope $\alpha=-1.3$, and high-mass (>0.5\msun) slope $\alpha=-2.35$, with upper mass limit 100\msun\ \citep[see Section 3 in][]{Olivier2022}.

\citet{Olivier2022} found that the ionizing spectra of J104457 and J141851 are consistent with extremely metal-poor stellar populations as represented by this extreme model that requires the addition of a hard photoionizing source, represented by this 80~kK blackbody, in order to produce the very-high-ionization lines. The physical origin of this hard photoionizing source is unknown, but could be a result of high-mass binaries, supersoft X-ray sources, stripped stars and Wolf-Rayet stars, or a combination of these or other exotic populations (see \citealt{Olivier2022} for a full discussion).

The two models from galaxies J104457 and J141851 take the derived gas-phase metallicities of 5.8\% and 8.7\% solar, respectively. They assume a spherical cloud and scale nitrogen and carbon in the gas by their measured abundances (see \citealt{Olivier2022} for more details).

\subsection{Observations at $z\sim1.6$}
The \cite{Cleri2023} sample of \nev-emitting galaxies is selected from the CLEAR \citep{Estrada-Carpenter2019,Estrada-Carpenter2020,Simons2021,Simons2023} parent catalog of \hst/Wide Field Camera 3 (WFC3) G102 and G141 grism observations using the following selection criteria:
\begin{itemize}
    \item Require a grism spectroscopic redshift, $1.39 < z < 2.30$, such that both \nev\ and \oiii\ are within the observed-frame spectral range of G102 and G141 sensitivity (0.8-1.65~$\mu$m).
    \item Require signal-to-noise ratio (\snr) of at least 3 for $\nev\ \lambda 3427$ and $\oiii\ \lambda\lambda 4960,5008$.
    \item Visual inspection of direct images with 1D and 2D spectra to ensure that no objects with poor continuum modeling and/or bad contamination subtraction make it into the final selection.
\end{itemize}
This selection leaves 25 galaxies with significant \nev\ emission. See \cite{Cleri2023} for further details on the individual steps of the sample selection.

\subsection{Observations at $5<z<8.5$}
To test our models against galaxies in the epoch of reionization, we use the \cite{Trump2023} sample. This sample contains five galaxies at $5.2<z<8.5$ from \jwst/NIRSpec observations of the SMACS 0723 Early Release Observations \citep{Pontoppidan2022}. Each galaxy was observed with NIRSpec in two visits (s007 and s008, see \citealt{Trump2023}). 

\begin{figure*}[t]
\epsscale{1.1}
\plotone{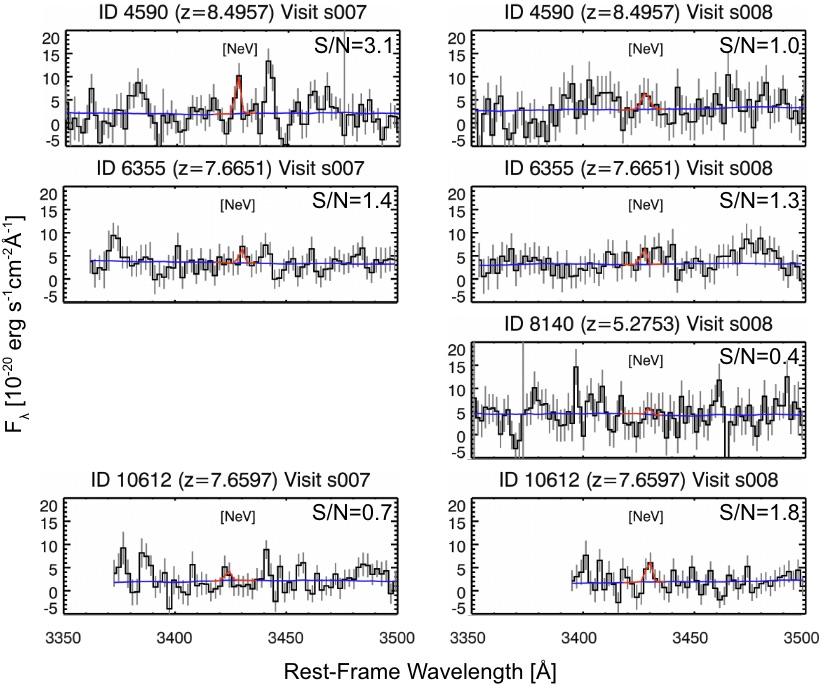}
\caption{Fits to \xneva\ for the four $5.3 < z < 8.5$ SMACS ERO galaxies from the \cite{Trump2023} sample with \nev\ coverage in \jwst/NIRSpec. Each row shows the spectra for one object (with IDs from \citealt{Trump2023} labeled) where the two panels in each row show the spectra from the two different Visits (```s007'' and ``s008'', respectively). The blue and red lines denote the continuum and emission line fits, respectively. We show the signal-to-noise ratio (\snr) for the fits in each visit. We do not show the first visit for ID 8140 as there is a spectral artifact at the observed wavelength of \xneva, making an accurate line measurement impossible. We note that none of these objects are well-detected (\snr\ > 3) in \nev\ by coadded spectra between visits. The \snr\ ratios and relevant emission-line ratios for these objects are given in Table \ref{tab:snr}.
\label{fig:smacs_spectra}}
\end{figure*}

The \cite{Trump2023} results include measurements for several emission lines of interest: $\oii\ \lambda\lambda 3727,3729$, $\neiii\ \lambda 3870$, $\oiii\ \lambda 4364$, \Hg, $\heii\ \lambda 4687$, \Hb, $\oiii\ \lambda 4960$, and $\oiii\ \lambda 5008$. 
In this work, we have refit the galaxies from the\ \citet{Trump2023} sample to probe for emission from  $\nev\ \lambda 3427$.  In three of the five sources we find possible detections of $\nev$ with \snr\ 1-3.  Of the two remaining galaxies, one shows no evidence of  \nev\ emission (\snr<1). In the remaining galaxy, (ID 5144 in the catalog of \citeauthor{Trump2023}), the $\nev\ \lambda 3427$ line is not covered as its wavelength falls in a detector gap in the NIRSpec gratings. Figure \ref{fig:smacs_spectra} shows fits to the emission in the region of  \nev\ for the four galaxies with coverage. The detections of \nev\ are marginal (1$\lesssim$\snr$\lesssim$ 3) at best, but we include them as useful upper limits in our analysis below.  We report the \snr\ ratios of the relevant emission lines as well as the \nev/\neiii\ limits (see Section \ref{subsec:ne53}), \neiii/\oii, and \oiii/\hb\ ratios and their uncertainties in Table \ref{tab:snr}. All of the values presented in Table \ref{tab:snr} are for the coadded spectra between the two visits. \xneva\ cannot be measured for ID~5144 as \nev\ lies in a detector gap at this redshift. There is also a spectral artifact in the region of \xneva\ in one visit of ID 8140.

\begin{deluxetable*}{lr|cccccccc}[t]
\tablecaption{Measurements from NIRSpec data on SMACS Galaxies} \label{tab:snr}
\tablehead{
\colhead{ID} & \colhead{Redshift} & \colhead{\xneva}\tablenotemark{i} & \colhead{\xoii}\tablenotemark{i} & \colhead{\xneiii}\tablenotemark{i} & \colhead{\Hb}\tablenotemark{i} & \colhead{\xoiiia}\tablenotemark{i} & \colhead{\nev/\neiii}\tablenotemark{ii} & \colhead{\neiii/\oii} & \colhead{$\oiii\lambda5008/\hb$}
}
\startdata
 4590 & 8.4957 & 2.3 & 3.2 & 6.8 & 20.4 & 38.3 & <0.61 & $1.82 \pm 0.63$ & $3.05 \pm 0.17$   \\
 5144 & 6.3792 & \dots & 7.4 & 10.8 & 27.4 & 86.3 & \dots & $1.32 \pm 0.22$ & $6.45 \pm 0.25$   \\
 6355 & 7.6651 & 1.8 & 23.5 & 15.5 & 26.3 & 96.5 & <0.20 & $0.48 \pm 0.04$ & $8.23 \pm 0.32$   \\
 8140 & 5.2753 & 0.4 & 13.7 & 4.6 & 6.9 & 28.6 & <0.44 & $0.39 \pm 0.09$ & $6.82 \pm 0.98$   \\
10612 & 7.6597 & 1.8 & 4.2 & 13.0 & 23.6 & 77.5  & <0.36 & $1.84 \pm 0.48$ & $6.97 \pm 0.31$ 
\enddata
\tablenotetext{i}{\snr\ ratio of the respective emission line.}
\tablenotetext{ii}{3$\sigma$ upper limit for \xneva\ nondetections. }

\end{deluxetable*}

\section{Emission-Line Ratio Classifications of Stellar Populations and AGN}\label{sec:sf_agn}

\subsection{The OHNO Diagram}

The OHNO (\oiii/\Hb\ and \neiii/\oii) diagram \citep{Zeimann2015,Backhaus2022a} is an emission-line ratio diagnostic designed to separate star-forming galaxies from AGN around $z\sim2$. This diagnostic compares ratios of emission lines at similar wavelengths (\xoii, \xneiii, \hb,  and \xoiii) where the production of \oiii\ and \neiii\ both require higher photon energies: the first two ionization energies of oxygen (the minimum energies required to produce \oii\ and \oiii\ photons) are 13.62 eV and 35.12 eV, and the second ionization energy of neon (the minimum energy required to produce \neiii\ photons) is 40.96 eV. Galaxies with strong \oiii/\Hb\ and/or \neiii/\oii\ require harder radiation fields, typically found in the emission-line regions of AGN.  \cite{Backhaus2022a} showed that division in the OHNO line ratios separates X-ray--selected AGN from non-AGN (based on classifications from the deep X-ray data in the Chandra Deep Fields at $z\sim1$). \cite{Backhaus2022a} defines the OHNO star-formation/AGN dividing line as 
\begin{equation}\label{eq:ohno}
\log\left(\frac{\oiii}{\Hb}\right)=\frac{0.35}{2.8\log( \neiii/\oii) - 0.8} + 0.64
\end{equation}

The OHNO diagnostic has also been used to show that \nev-emitting galaxies near the peak of cosmic star formation and AGN activity at $z\sim 1-2$ are preferentially classified as AGN \citep{Backhaus2022a,Cleri2023}. 

\begin{figure*}[t]
\epsscale{1.1}
\plotone{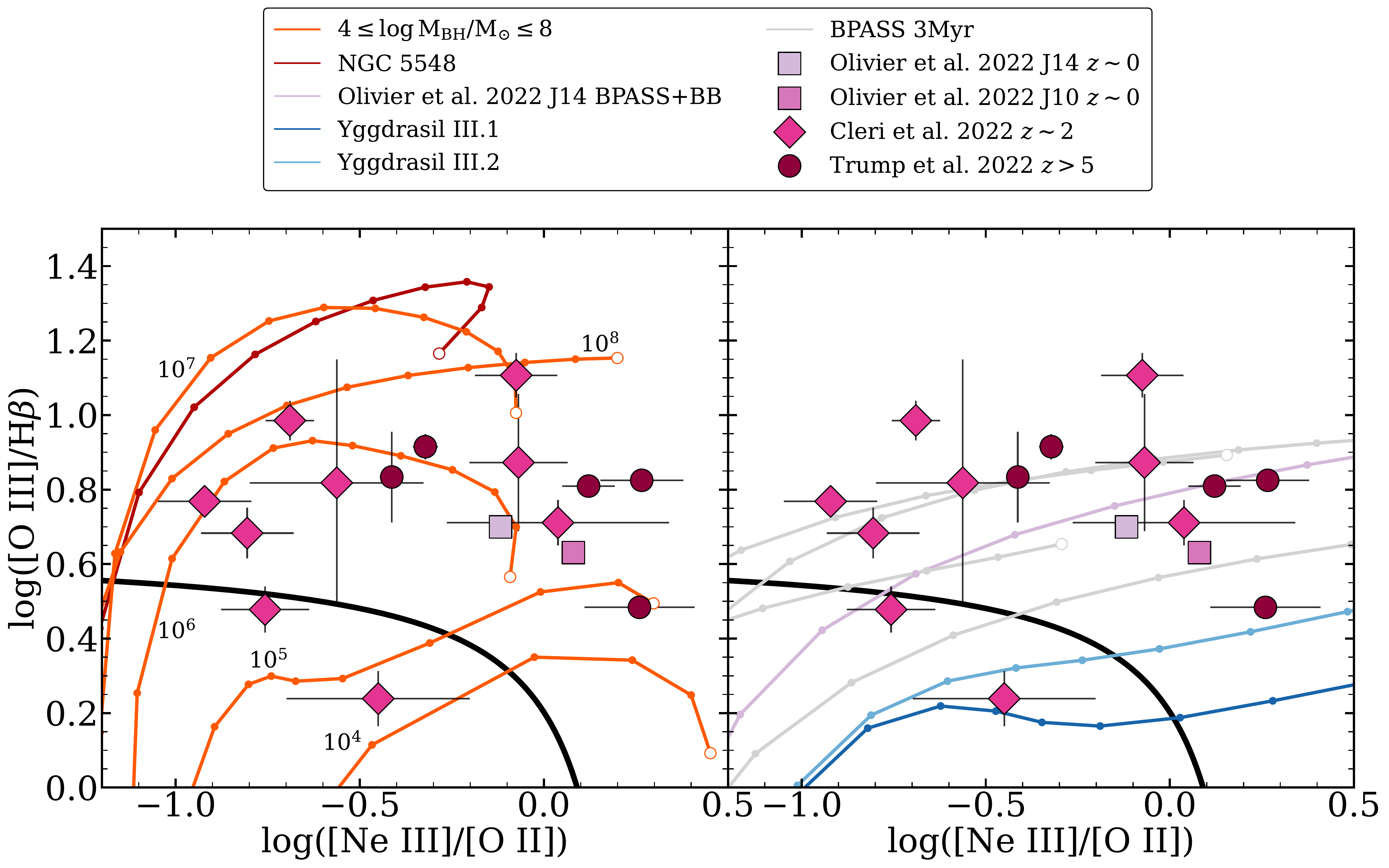}
\caption{
The OHNO diagram of log(\oiii/\Hb) versus log(\neiii/\oii). Models are shown in a grid of dimensionless ionization parameter $-4\leq\log U\leq-1$, with the open symbol at the end of each track marking $\log U =-1$. \textit{Left:} AGN models in OHNO parameter space. We also show the NGC 5548 SED from \citealt{Mehdipour2015} in red, and black hole accretion disk models from $4\leq\log\mbh/\msun\leq8$ in orange, with annotations indicate the black hole mass of each model. \textit{Right:} Stellar population models in OHNO parameter space. We show the BPASS 3 Myr stellar populations of a range of metallicities $Z=0.00005, 0.05, 0.10, 0.20, 0.50, 1Z_\odot$ in gray. We also include the \cite{Olivier2022} BPASS+blackbody model for J141851 in purple, and the Yggdrasil Population III.1 and III.2 models in dark and light blue, respectively. In each panel we show the $z\sim0$ data from \citealt{Olivier2022}, $z\sim2$ observations from \citealt{Cleri2023}, and $z>5$ observations from \cite{Trump2023}. 
\label{fig:ohno}}
\end{figure*}

At high redshifts, it becomes difficult to distinguish between galaxies with photoionization from stellar populations and those with AGN using only OHNO.   We show the OHNO diagram in Figure \ref{fig:ohno}. We also show our stellar, stellar+blackbody, and AGN models. We show our $z\sim0$, $z\sim1.6$, and $5<z<8.5$ comparison samples. We note that nearly all of the \nev-emitting objects and the SMACS objects with \nev\ limits are consistent with photoionization dominated by AGN activity by the \citealt{Backhaus2022a} diagnostic. For maximal completeness, we require a \snr\ greater than 1 for the OHNO emission lines. The black hole accretion models and the NGC 5548 model show line ratios consistent with expectations for traditional AGN at high ionization parameters. We also show that extreme stellar populations are capable of producing line ratios which would be labeled as AGN by the \citealt{Backhaus2022a} diagnostic.

\subsection{The Ne53 Ratio}\label{subsec:ne53}
As \nev\ requires ionization energies much higher than those of most other strong rest-frame UV/optical emission lines (97.11-125.26 eV), it probes extremely energetic photoionization missed by lower-energy tracers. By taking the ratio of \nev\ to lower ionization species, we can trace physical conditions of the ISM missed by other line ratios, such as \oiii/\hb, \neiii/\oii, \sii/\ha\ or \nii/\ha. We define the Ne53 ratio as 
\begin{align}\label{eq:ne53}
    \mathrm{Ne53} &= \frac{\xneva}{\xneiii}
\end{align}

Ne53 is an opportunistic line ratio of study for several reasons. These species of neon evolve strongly with temperature and are relatively insensitive to changes in density (see Appendix A of \citealt{Cleri2023}), both \nev\ and \neiii\ are accessible to \jwst/NIRSpec spectroscopy from $z \sim 2$ to 12, and are close enough in wavelength for the ratio to be relatively insensitive to dust attenuation and instrumental effects such as calibration uncertainties. These two lines in particular also have the advantage of not being blended with other significant spectral features. For this reason, we choose to not use the $\nev~\lambda 3347$, $\neiii ~\lambda 3968$, or $\neiii~\lambda 3343$ lines in this ratio due to blending with each other or nearby lines at lower spectral resolutions. 

The advantage of Ne53 is the large difference in minimum energy required to produce the emission lines: 97.11 eV for \nev\ and 40.96 eV for \neiii. This traces the ``very high'' to ``high'' ionization zones of \cite{Berg2021}. Ne53 also has the advantage of eliminating any abundance degeneracy from using spectral lines of two different elements.  

\begin{figure*}[t]
\epsscale{1.1}
\plotone{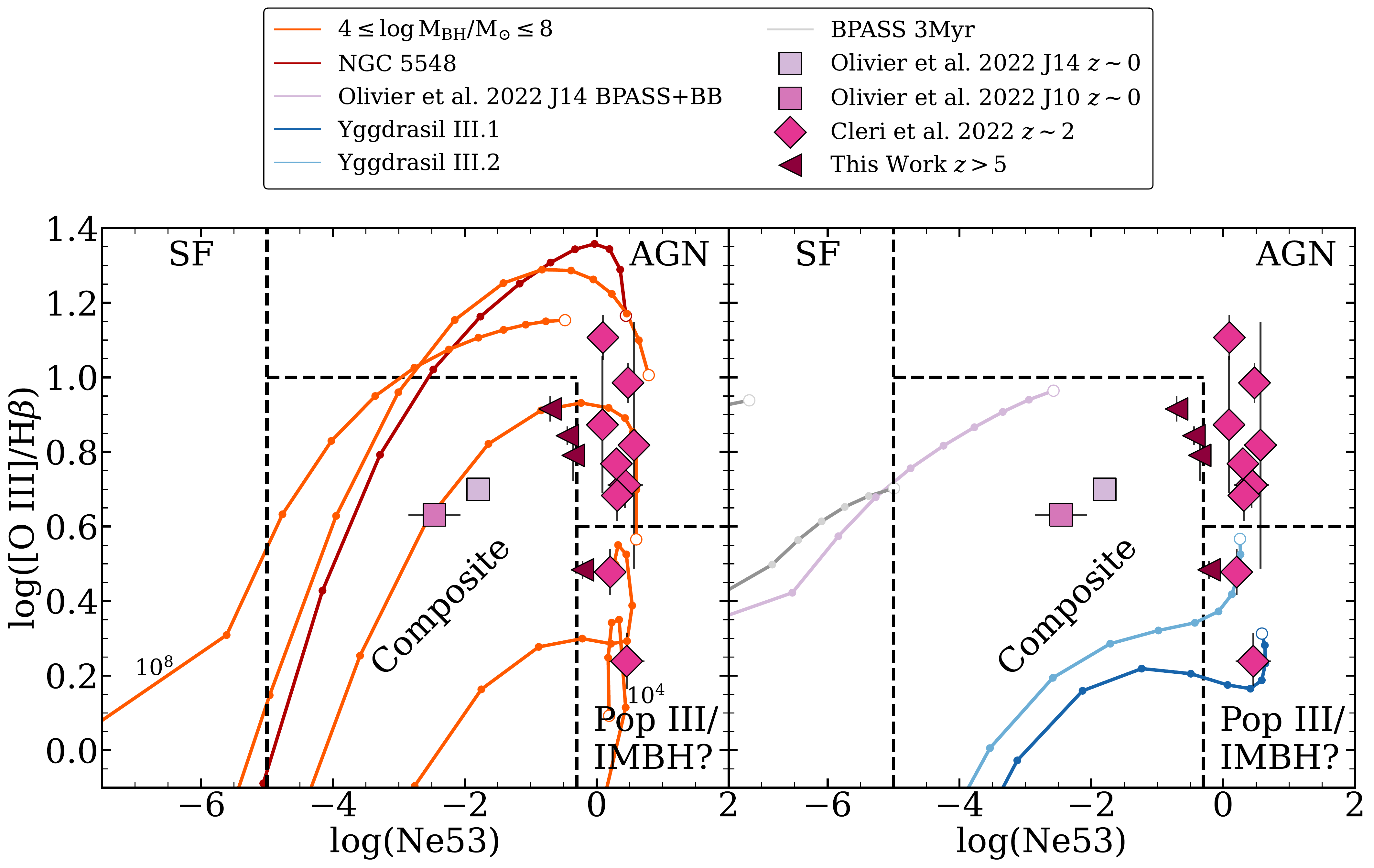}
\caption{
The relation between \oiii/\hb\ and \nev/\neiii\ (Ne53) ratios. Models are shown in a grid of dimensionless ionization parameter $-4\leq\log U\leq-1$, with the open symbol at the end of each track marking $\log U =-1$. \textit{Left:} log(\oiii/\Hb) versus log(Ne53) showing AGN models. We show the variable \mbh\ accretion disk models in orange, marking the  extremes of $\log\mbh/\msun = 8$ and $\log\mbh/\msun = 4$, and the NGC 5548 SED from \citealt{Mehdipour2015} in red. \textit{Right:} log(\oiii/\Hb) versus log(Ne53) showing stellar population models. We show the BPASS 3 Myr stellar population models in gray for metallicities $Z=0.00005, 0.05, 0.10, 0.20, 0.50, 1Z_\odot$ (most lines are off the plot left). We also include the \cite{Olivier2022} BPASS+blackbody model for the local high-ionization metal-poor dwarf J141851 in purple, and the Yggdrasil Population III.1 and III.2 models in dark and light blue, respectively. In each panel we show the $z\sim0$ data from \citealt{Olivier2022}, $z\sim2$ observations from \citealt{Cleri2023}, and $z>5$ observations from \cite{Trump2023} including their 3$\sigma$ upper limits on Ne53 as measured in this work. We note a distinct separation between the parameter space covered by the AGN models and the BPASS stellar populations. The BPASS models are unable to reproduce observed Ne53 ratios, even at the highest ionization parameters. The dashed black lines delineate regions where the line ratios are mostly consistent with ionization from  star-formation (SF), AGN, a composite of potential SF and AGN, and possible ionization from Pop III stars or accretion onto IMBHs.
\label{fig:ne53}}
\end{figure*}

We explore Ne53 as an indicator of black hole accretion or of the hard ionizing radiation from extreme stellar populations in Figure \ref{fig:ne53}. We show a clear difference of several orders of magnitude between the AGN and BPASS/BPASS + blackbody models in Ne53 space; especially interesting is that the Pop III star models are competitive with AGN in the production of \nev\ relative to \neiii, with the Pop III models residing a similar region of the parameters space as the low-mass ($\mbh\leq10^5\msun$) black hole accretion disk models. The BPASS-only models fail to produce any meaningful \nev, with log(Ne53) ratios peaking around -5 for all metallicities. 

We include polygons in Figure \ref{fig:ne53} to mark the general locations of different ionizing engines in this parameter space. We empirically separate the \oiii/\hb\ versus Ne53 diagram into four regions:
\begin{itemize}
    \item logNe53 $<-5$: Objects dominated by ``normal''  star formation alone. 
    \item log\oiii/\hb\ $<$ 1 AND $-5<$ logNe53 $<-0.3$: ``Composite'' zone of multiple potential ionizing sources. 
    \item (log\oiii/\hb\ $>1$ AND $-5<$ logNe53 $<-0.3$) OR (log\oiii/\hb\ $>0.6$ AND logNe53 $>-0.3$): ``Traditional'' SMBH accretion disk AGN. 
    \item log\oiii/\hb\ $<0.6$ AND logNe53 $>-0.3$: Pop III stars or IMBH accretion disks.
\end{itemize}

We show our photoionization models compared to our $z\sim0$, $z\sim1.6$, and $5<z<8.5$ samples to show that objects with well-detected \nev\ emission are many orders of magnitude inconsistent with stellar population-only models from BPASS, even at the highest ionization parameters. We also show that the 3$\sigma$ upper limits of Ne53 place the SMACS galaxies at $5<z<8.5$ near the boundary of the Composite and AGN/Pop III/IMBH regions, but we do not draw any significant conclusions to to the low \snr\ of these measurements. 

\section{Discussion}\label{sec:discussion}
The emission-line ratio diagnostics of Figures \ref{fig:ohno} and \ref{fig:ne53} carry many implications of the nature of EIGs and \nev\ emission. The Ne53 ratio (see Figure~\ref{fig:ne53})  separates AGN and Pop III models from BPASS/BPASS + blackbody stellar models by several orders of magnitude at a given ionization parameter. As such, we conclude that the Ne53 ratio is an effective diagnostic of hard radiation fields.  However, there is a large region where the Ne53 and \oiii/\hb\ ratios could stem from star formation or AGN (labeled ``composite'' in Figure \ref{fig:ne53}), but any object that resides in this region is of interest as they will be EIGs with possible indications of exotic stellar populations, accreting IMBH, or both.   Therefore, Ne53 alone is not sufficient to separate AGN from extreme stellar populations, such as those presented in our Pop III models. 

The OHNO diagnostic (Figure~\ref{fig:ohno}) shows that nearly all of the compared galaxies with \nev\ detections (or upper limits to \nev\ for the SMACS objects) are consistent with photoionization from an AGN, as defined by the \citealt{Backhaus2022a} division (Equation \ref{eq:ohno}). This result shows that the OHNO diagram is not an accurate diagnostic for the most extreme emission-line galaxies in the local Universe, given our comparison to J141851 and J104457 from \citealt{Olivier2022} and \citealt{Berg2019,Berg2021}. This is likely due to the calibration of the OHNO AGN/SF division from \citealt{Backhaus2022a} being based on much higher mass X-ray AGN at $z\sim 1-2$, and as such OHNO is not designed to discriminate AGN emission from low-mass metal-poor EIGs like J104457 and J141851. 

If very high redshift galaxies are similar to the EIGs studied by \citealt{Olivier2022}, then the \citep{Backhaus2022a} OHNO division between AGN and star-forming galaxies may be inaccurate. As such, we do not adhere to the OHNO classifications of the SMACS ERO objects at $z>5$ as AGN. Other works have conjectured that a subset of these objects may be narrow-line AGN \citep{Brinchmann2023}, but there is not conclusive evidence from other metrics for the AGN nature of any of these galaxies.

Our black hole accretion disk models for AGN cover a wide region of the OHNO parameter space. The higher black hole mass ($\mbh\geq10^6\msun$) models produce higher \oiii/\hb\ and lower \neiii/\oii\ ratios than the lower black hole mass models. The higher black hole mass models are similar to the galaxies in the $z\sim2$ comparison sample of \nev\ emitting galaxies from \citealt{Cleri2023}, which is most similar to the \citealt{Backhaus2022a} sample from which the OHNO diagnostic was derived. 

The photoionization models show increased \neiii/\oii\ compared to \oiii/\hb\ for AGN, along with increased \oiii/\hb\ from the BPASS+blackbody models from \cite{Olivier2022}. As a useful proof of concept for the variable \mbh\ accretion disk models, we note that the NGC 5548 model resides in a similar parameter space as the high-mass ($\mbh\geq10^7\msun$) \cite{Done2012} models. We also note that the Pop III models occupy the lowest \oiii/\hb\ but highest \neiii/\oii\ of any of the models presented.  

Other non-AGN and non-SF explanations of \nev\ emission remain an open question. Other works have attributed \nev\ emission in small samples of galaxies to shocks from supernovae \citep{Izotov2012,Izotov2021,Leung2021}, but are unable to rule out production via AGN on a population level.

Perhaps the most intriguing alternative explanation for \nev\ production in the early Universe is through exotic Pop III stars. While exact constraints on the nature of Pop III stars and the high-redshift IMF remain unknown, we show that current models of Pop III stars are indeed capable of providing the energetic radiation necessary to produce ultra-high ionization emission features such as \nev\ (see Figure \ref{fig:ne53}). 

Throughout the analyses performed in this work, we have identified an interesting diagnostic for Pop III stars and IMBH accretion disks. The emission line ratio relations in Figures \ref{fig:ohno} and \ref{fig:ne53} can be used in conjunction to separate objects four distinct groups (see Section \ref{sec:sf_agn} for exact delineations): 
\begin{itemize}
    \item Objects dominated by star formation do not produce significant \nev, thus lie very low (or undetectable) in Ne53 (Figure \ref{fig:ne53}). 
    \item AGN dominated by supermassive ($\mbh\geq10^6\msun$) black hole accretion disks produce high \oiii/\hb\ ratios, consistent with traditional star formation/AGN diagnostics \citep[e.g.,][]{Veilleux1987,Baldwin1981,Backhaus2022a}.
    \item Objects dominated by Pop III stars or AGN with lower-mass black holes (IMBHs, with $\mbh\lesssim10^5\msun$)  reside in a unique parameter space in the emission line ratio relations in this work. These objects are low in \oiii/\hb, yet high in both \neiii/\oii\ and Ne53, similar to the parameter space occupied by the Pop III models. 
    \item A composite region for objects which have spectra with contributions from a variety of ionizing sources. 
\end{itemize}

Objects with significant contribution to their spectra from Pop III stars and lower mass black hole accretion disks are both poorly understood. Information from emission-line ratios such as those presented in this work will offer insight into the physical conditions of these interesting objects. 

We look to future observations from the \jwstlong\ and next-generation facilities to offer constraints on these exotic stellar populations through emission-line spectroscopy of EIGs. Recent studies \citep{Trussler2022} have suggested that \jwst\ is capable of detecting $M_*=10^6~\msun$ Pop III galaxies at $z\sim 8$ spectroscopically, with (very) deep integrations of tens to hundreds of hours of NIRSpec/G140M time. However, \cite{Trussler2022} calculates that moderately-lensed or more massive ($M_*=2-3\times10^6~\msun$) Pop III galaxies will be detectable with medium sized \jwst\ General Observer programs (25-75 hours). 

We caution against the general use of lower-ionization emission-line ratio diagnostics of star-formation and AGN (e.g., BPT, VO87, OHNO, etc.) for high-redshift systems as the sole discriminator between sources of ionization. In general, the use of emission-line ratio diagnostics for objects or redshift regimes for which they are not designed is questionable. Recent works have discussed the need for more emission-line ratio diagnostics which are more robust across redshifts, e.g., Figure \ref{fig:ne53} in this work. Instead, we suggest the aggregation of multiple tracers of black hole accretion/extreme stellar populations to make decisive conclusions about the nature of the ionizing spectra of high-redshift objects: emission-line velocity profiles, broad line components, multiwavelength luminosities (X-ray, radio, mid-IR, ...), etc. (see \citealt{Larson2023} for a discussion of AGN tracers for the current highest-redshift AGN candidate at z=8.7).

\section{Summary and Conclusions}\label{sec:conclusions}
In this work, we used photoionization models to explore the physical conditions of \xneva\ emission in extreme ionization galaxies (EIGs) across cosmic time. We produce \cloudy\ photoionization models using incident SEDs from  BPASS stellar populations alone \cite{Stanway2018}, variable mass black hole accretion disks \citep{Done2012,Cann2018}, the well-studied local Seyfert NGC 5548 \cite{Mehdipour2015}, and the Yggdrasil Population III.1 and III.2 models \citep{Zackrisson2011}. We also include the BPASS stellar populations plus an 80 kK blackbody model of local metal-poor dwarf galaxy J141851 \citep{Olivier2022}. We compare our results to observations of galaxies at $z\sim 0$ \citep{Berg2016,Berg2019,Berg2021,Olivier2022}, $z\sim 2$ \citep{Cleri2023}, and $5>z>8.55$ \citep{Trump2023}.

The primary findings of this work are as follows:
\begin{itemize}
    \item We fit for \nev\ in the $z>5$ SMACS 0723 sample from \citealt{Trump2023}, and find no significant (\snr\ > 3) detections in the four objects for which there is \jwst/NIRSpec coverage of \nev. We find marginal (1<\snr<3) detections in three objects (one object with \snr<1), which we use to assess the limiting behavior in our emission-line ratio diagnostics. We publish our measured \xneva\ \snr\ ratios and emission-line ratios in Table \ref{tab:snr}.
    \item The \nev/\neiii\ (Ne53) ratio is an effective indicator of photoionization from a source other than ``normal'' stellar populations. Current photoionization models cannot reproduce \nev\ emission with ``normal'' stellar populations alone  (such as those from the BPASS models, see Figure \ref{fig:ne53}). We conclude that the measurement of any meaningful detection of Ne53 (log Ne53 $\gtrsim$ $-$5) likely implies the presence of an ionizing engine other than ``normal'' stellar populations alone (e.g., likely either AGN or Population III stars). 
    \item We use Ne53 in conjunction with other line ratios to produce a diagnostic of AGN and Population III star activity. We propose the combination of Figures \ref{fig:ne53} and \ref{fig:ohno} to be used as such a diagnostic, where we anticipate galaxies with significant contributions of their ionizing radiation from Population III stars or low-mass AGN ($\mbh\leq10^5\msun$) to reside in low \oiii/\hb\ (log \oiii/\hb\ $\lesssim$ 0.6) and high \neiii/\oii\ (log \neiii/\oii\ $\gtrsim$ 0) and Ne53 (log Ne53 $\gtrsim$ -0.3, indicating very high ionization) parameter spaces. We propose that Figure \ref{fig:ne53} can be used in future studies in conjunction with other AGN tracers (e.g., velocity profiles, broad line components, X-ray/radio/mid-IR luminosities) to further separate sources of ionizing radiation.
\end{itemize}

Our results show that \nev\ emission probes highly energetic photoionization ($\sim$100 eV), which, for the models used in this work, is only reproducible by black hole accretion or Pop III stars. Although we attribute the \nev\ production in many of the galaxies in our comparison samples, we also note that there are other production mechanisms, namely supernova shocks or other exotic stellar populations (e.g., Wolf-Rayet stars). We note that the unknown nature of Pop III stars and low-mass black hole accretion disks in the early Universe indicates that epoch of reionization galaxies are an exciting laboratory in which to study these extreme emission lines. We also note that the potential coevolution of accreting black holes and stellar populations in the early Universe can have competing contributions to the production of high-ionization emission lines. Regardless, galaxies that show strong Ne53 and low \oiii/\hb\ will be candidates for rare sources and may guide us to galaxies with ionization from either the first stars or AGN with lower mass black holes.  

As such, these results motivate future observations of very high-ionization emission lines, like $\nev ~\lambda 3427 \AA$, using cutting-edge observatories like the \jwstlong. \jwst\ will reach a flux limit that is an order of magnitude fainter than the \hst\ data from the \cite{Cleri2023} sample for similar exposure times, enabling detection of fainter \nev-line emission. \jwst\ is outfitted with NIRSpec and NIRCam, both of which had spectroscopic capabilities covering strong  UV high-ionization emission lines, like the \nev\ doublet, over a range of redshift (e.g., $6<z<11$). Even \jwst/NIRISS can cover of \nev\ at slightly lower redshift ranges ($3<z<7$). 

\jwst/MIRI even provides low (R$\sim$100) and medium resolution (R$\sim$1500-3500) spectroscopy, which covers 5--28 $\mu$m.  This will be able to observe rest-frame UV/optical spectral features for bright galaxies out to ultra-high-redshifts ($z>15$). Spectroscopy in this very early epoch in cosmic history will give direct measurements of first-generation stellar populations, as well as the beginning of chemical enrichment of the ISM. 

The potential degeneracy of \nev\ production from low-mass black hole accretion disks or Pop III stars in the early Universe is difficult to break with rest-UV/optical line ratios alone. In addition to the integrated emission-line diagnostics we present in this work (Figures \ref{fig:ohno} and \ref{fig:ne53}), we can pose a solution using the integral field unit (IFU) capabilities of \jwst/NIRSpec by looking at the spatial distribution and compactness of the high-ionization emission. We expect galaxies with significant Pop III fractions to be more uniformly distributed in their \nev\ emission (or other very high ionization emission) than an AGN, where we expect the emission to be highly centrally dense. We can also break this degeneracy with more spectral information, including more detailed chemical abundances and metallicities. 

Future studies using current and next-generation systems of high-redshift systems through the lens of extreme high-ionization emission lines, including \nev, will offer long-awaited answers to the underlying physics of the epoch of reionization. 


\software{\texttt{grizli} \citep{Brammer2008}, FAST \citep{Kriek2009}, EAZY \citep[]{Brammer2008, Wuyts2011}, Astropy \citep{Astropy2013}, NumPy \cite{Harris2020}, Matplotlib \citep{Hunter2007}}, \cloudy\ \citep{Ferland2017}, seaborn \citep{Waskom2021}, pandas \citep{Reback2022}, \texttt{XSPEC} \citep{Arnaud1996}

\acknowledgements 

This work is based on data obtained from the Hubble Space Telescope through program number GO-14227. Support for Program number GO-14227 was provided by NASA through a grant from the Space Telescope Science Institute, which is operated by the Association of Universities for Research in Astronomy, Incorporated, under NASA contract NAS5-26555. NJC, JRT, and BEB acknowledge support from NSF grant CAREER-1945546 and NASA grants JWST-ERS-01345 and 18-2ADAP18-0177.  NJC, JRT, CP and BEB acknowledge support NASA grant JWST-ERS-01345. NJC and CP also acknowledge support from NASA/\textit{HST} AR 16609.  This work acknowledges support from the NASA/ESA/CSA James Webb Space Telescope through the Space Telescope Science Institute, which is operated by the Association of Universities for Research in Astronomy, Incorporated, under NASA contract NAS5-03127. Support for program No. JWST-ERS01345 was provided through a grant from the STScI under NASA contract NAS5-03127. TAH and AY are supported by appointments to the NASA Postdoctoral Program (NPP) at NASA Goddard Space Flight Center, administered by Oak Ridge Associated Universities under contract with NASA.

The authors thank Shobita Satyapal, Jeffrey Mckaig, and Jenna Cann for discussion and providing assistance with photoionization modeling. NJC thanks Jonathan Cohn, Justin Cole, and Maeve Curliss for insightful scientific and data visualization discussions. NJC also thanks the CANDELS Lyman-$\alpha$ Emission at Reionization (CLEAR) and Cosmic Evolution Early Release Science (CEERS) collaborations for influential discussions.  

Some of the data presented in this paper were obtained from the Mikulski Archive for Space Telescopes (MAST) at the Space Telescope Science Institute. The \cite{Berg2019,Olivier2022} data can be accessed via \dataset[DOI: 10.3847/1538-4357/ac8f2c]{https://doi.org/10.3847/1538-4357/ac8f2c}. The CLEAR \citep{Simons2023} data can be accessed via \dataset[DOI: 10.3847/1538-4365/acc517]{https://doi.org/10.3847/1538-4365/acc517}. The SMACS 0723 data from \cite{Trump2023}  can be accessed via \dataset[DOI: 10.3847/1538-4357/acba8a]{https://doi.org/10.3847/1538-4357/acba8a}.

\clearpage
\bibliography{library}{}

\end{document}